\documentclass[letterpaper,twocolumn,10pt]{article}
\usepackage{usenix}

\usepackage{amsmath,amssymb,amsfonts}
\usepackage{algorithm}
\usepackage{algpseudocode}
\usepackage{graphicx}
\usepackage{multirow}
\usepackage{array}
\usepackage{subfigure}
\usepackage{tabularx}
\usepackage{siunitx}
\usepackage{textcomp}
\usepackage{adjustbox} 
\usepackage{colortbl} 
\usepackage{xcolor}
\usepackage{booktabs}
\usepackage{tcolorbox}

\begin{document}

\newcommand{\carrow}{\textcolor{blue}{$\rightarrow$}}
\definecolor{red}{HTML}{cd5946} 
\definecolor{green}{HTML}{4c9e66} 

\date{}

\title{\Large \bf VulnBot: Autonomous Penetration Testing for A Multi-Agent Collaborative Framework}

\author{
  \rm He Kong\textsuperscript{1,2}, 
  \rm Die Hu\textsuperscript{1,2}, 
  \rm Jingguo Ge\textsuperscript{1,2}, 
  \rm Liangxiong Li\textsuperscript{1}, 
  \rm Tong Li\textsuperscript{1} , and 
  \rm Bingzhen Wu\textsuperscript{1} \\
  \textsuperscript{1}State Key Laboratory of Cyberspace Security Defense, Institute of Information Engineering, \\Chinese Academy of Sciences \\
  \textsuperscript{2}School of Cyber Security, University of Chinese Academy of Sciences
}
\maketitle

\begin{abstract}
Penetration testing is a vital practice for identifying and mitigating vulnerabilities in cybersecurity systems, but its manual execution is labor-intensive and time-consuming. Existing large language model (LLM)-assisted or automated penetration testing approaches often suffer from inefficiencies, such as a lack of contextual understanding and excessive, unstructured data generation. 
This paper presents VulnBot, an automated penetration testing framework that leverages LLMs to simulate the collaborative workflow of human penetration testing teams through a multi-agent system. To address the inefficiencies and reliance on manual intervention in traditional penetration testing methods, VulnBot decomposes complex tasks into three specialized phases: reconnaissance, scanning, and exploitation. These phases are guided by a penetration task graph (PTG) to ensure logical task execution. Key design features include role specialization, penetration path planning, inter-agent communication, and generative penetration behavior. Experimental results demonstrate that VulnBot outperforms baseline models such as GPT-4 and Llama3 in automated penetration testing tasks, particularly showcasing its potential in fully autonomous testing on real-world machines.
\end{abstract}

%-------------------------------------------------------------------------------
\begin{figure*}[ht]
\centering
\includegraphics[width=\linewidth]{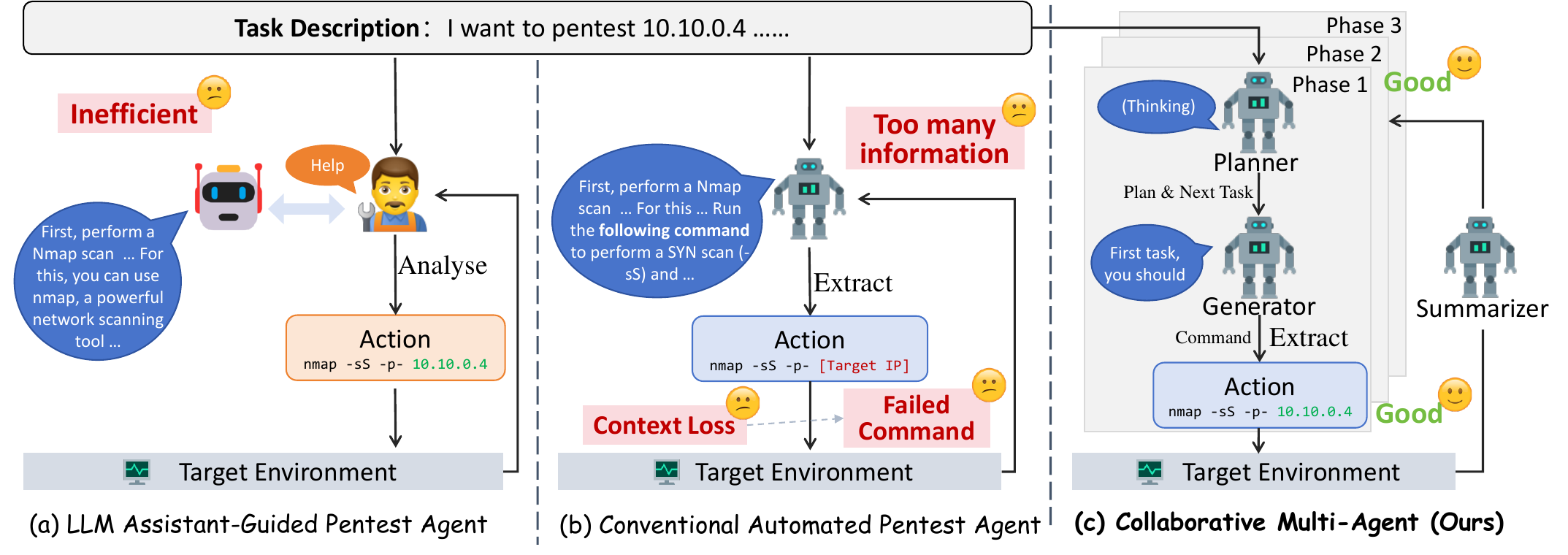}
\caption{The workflow comparison of three approaches to automated penetration testing: (a) LLM Assistant-Guided Pentest Agent, which requires assistance due to inefficiency; (b) Conventional Automated Pentest Agent, which struggles with information overload and context loss; and (c) Collaborative Multi-Agent system, which employs a phased and modular approach, enhancing the overall efficiency and autonomy of the penetration testing process through multi-agent coordination.}
\label{fig:motivation}
\end{figure*}
\section{Introduction}
Penetration testing is a critical methodology for proactively identifying network vulnerabilities and mitigating potential cyberattacks \cite{arkin2005software, fatima2023impact}. It enables the timely detection of weaknesses in target systems, facilitating targeted remediation and reinforcement efforts. According to market forecasts, the penetration testing market is projected to grow significantly, expanding from US \$1.92 billion in 2023 to US \$6.98 billion by 2032 \cite{cyphere}. 
Despite its importance, traditional penetration testing remains a labor-intensive and time-consuming process, requiring highly skilled professionals to execute complex workflows manually. As network threats continue to grow in both complexity and scale, there is an urgent need for more efficient, scalable, and automated penetration testing methodologies to reduce manual effort and enhance testing efficiency \cite{bishop2007penetration}.

Recent advancements in large language models (LLMs) and multi-agent systems have opened new avenues for automating penetration testing \cite{hong2023metagpt,park2023generative, brown2020language, achiam2023gpt, team2023gemini, liu2024deepseek, dubey2024llama}. Consequently, researchers have proposed various approaches to leverage LLMs for automated penetration testing. For instance, Deng et al. introduced PentestGPT, a pioneering effort to utilize LLMs for automating penetration testing \cite{deng2024pentestgpt}. PentestGPT addresses the issue of context loss in LLMs during penetration testing through three interconnected modules: a reasoning module, a generation module, and a parsing module. However, PentestGPT heavily relies on human intervention and cannot assess the extent of such involvement, resulting in limited agent autonomy.
In contrast, AutoAttacker, a novel method, focuses on automating the post-penetration phase of simulated network attacks (i.e., "keyboard-operated" attacks) \cite{xu2024autoattacker}. By employing a modular design, AutoAttacker leverages the planning, summarization, and code generation capabilities of LLMs, combined with tools like Metasploit, to demonstrate the efficacy of LLMs in isolated security tasks. Nevertheless, AutoAttacker primarily targets specific tasks rather than real-world environments. While existing studies have explored the use of LLMs for automated penetration testing, they are often limited in scope, focusing on specific tasks or relying on detailed vulnerability descriptions, which are difficult to apply in real-world scenarios. Additionally, these methods predominantly depend on the GPT-4 model \cite{achiam2023gpt}, making it challenging to execute complex tasks using open-source models.

In this paper, we present VulnBot, an autonomous, multi-agent penetration testing framework based on LLMs, designed to emulate the collaborative workflows of human penetration testing teams. By integrating specialized modules that focus on different phases of penetration testing, VulnBot aims to automate and streamline the process of identifying vulnerabilities in target systems. The framework incorporates distinct roles, such as reconnaissance, scanning, and exploitation, along with a Penetration Task Graph (PTG)-based approach to penetration path planning, inter-agent communication, and generative penetration behavior. These components work together to simulate a robust and comprehensive penetration testing workflow.

At the core of VulnBot's design is its ability to model the penetration testing process as a series of interdependent tasks, each contributing to the overarching goal of identifying and exploiting vulnerabilities in the target system. The PTG organizes these tasks, ensuring that each step is executed in the correct sequence and context. VulnBot enhances inter-agent communication through a Summarizer module, which acts as a bridge between different phases of penetration testing. By summarizing key task outcomes and transmitting them to subsequent roles, the Summarizer ensures that critical information is preserved and prioritized across stages. This targeted communication minimizes redundancy, ensures clarity, and optimizes the flow of information across agents, thereby maintaining the integrity and continuity of the penetration testing process.
Furthermore, VulnBot's Generator and Executor modules translate these tasks into tool-specific commands, simulate human-like interactions with the target system, and autonomously execute them, significantly reducing the need for continuous human oversight. This paper introduces three operational modes: automatic, semi-automatic, and human-involved. The experimental evaluation focuses on the automatic mode, as human involvement introduces subjectivity and variability that are difficult to quantify.

We evaluated VulnBot across two distinct benchmarks to assess its performance and real-world applicability. On the AUTOPENBENCH, VulnBot significantly outperformed baseline models, including GPT-4o, Llama3.3-70B, and Llama3.1-405B. Specifically, VulnBot-Llama3.1-405B achieved a completion rate of 30.3\%, compared to 9.09\% for Llama3.1-405B and 21.21\% for GPT-4o. Additionally, VulnBot demonstrated superior performance in the early stages of the test. By delaying the automation of penetration testing to later stages, VulnBot ensures that critical subtasks are executed with greater precision, thereby increasing the likelihood of completing the testing process. Ablation studies further confirmed the effectiveness of the various components within the framework.
On real-world machines using the AI-Pentest-Benchmark, VulnBot—when paired with Llama3.1-405B and DeepSeek-v3—surpassed other baseline models. 
When integrated with Retrieval Augmented Generation (RAG), VulnBot’s performance further improved. In real-world machines, VulnBot with RAG autonomously completed tasks end-to-end, a feat that GPT-4o and Llama3.1-405B, which relied on human intervention, could not achieve. These results highlight VulnBot’s potential for fully autonomous penetration testing.

The contributions of this work are as follows:

\begin{itemize}
    \item We introduce VulnBot, an autonomous penetration testing framework that leverages the capabilities of LLMs and multi-agent systems to automate complex penetration testing workflows. Inspired by the collaborative dynamics of human penetration testing teams, VulnBot employs a tri-phase design—reconnaissance, scanning, and exploitation. This design ensures that agents focus on specific tasks, minimizing information loss and enhancing efficiency.
    
    \item We propose a task-driven mechanism based on a PTG, which models tasks and their dependencies as a directed acyclic graph. The PTG ensures that tasks are executed in a logical and conflict-free order, providing a structured framework for tracking task progress and outcomes. This mechanism, combined with a Check and Reflection Mechanism, enables continuous improvement and adaptation of the plan based on task status, achieving effective error handling and feedback.
    
    \item By utilizing open-source models such as Llama3.3-70B, Llama3.1-405B, and DeepSeek-V3, we demonstrate the feasibility and effectiveness of leveraging open-source LLMs for automated penetration testing. Our experimental results show that VulnBot outperforms baseline models such as GPT-4 and Llama3, achieving a 69.05\% subtask completion rate and a 30.3\% overall completion rate on the AUTOPENBENCH. Additionally, VulnBot achieved the best performance on six real-world machines in the AI-Pentest-Benchmark. Through the integration of RAG, VulnBot successfully realized complete end-to-end penetration of real-world machines.

\end{itemize}

%-------------------------------------------------------------------------------
\section{Background \& Motivation}
\begin{table*}[h!]
\centering
\small % Reduce font size
\renewcommand{\arraystretch}{1.3} % Adjust row height for better readability
\begin{adjustbox}{width=\textwidth}
\begin{tabular}{l|c|c|c|c|c|c|c|c}
\toprule
\multirow{2}{*}{\textbf{Model}} & \multirow{2}{*}{\textbf{Phase}} & \multirow{2}{*}{\textbf{Failure Count}} & \multicolumn{6}{c}{\textbf{Cause of Failure}} \\ \cmidrule(lr){4-9}
                               &                                 &                                         & \textbf{Session Context Loss} & \textbf{False Output Interpretation} & \textbf{Failed Tool} & \textbf{Deadlock Operation} & \textbf{Failed Command Param} & \textbf{Other}  \\ 
\midrule
\multirow{3}{*}{Llama3.3-70B} & Reconnaissance & 28 & 18 (64.29\%) & 3 (10.71\%) & 2 (7.14\%) & 0 (0.00\%) & 4 (14.29\%) & 1 (3.57\%)  \\
                              & Scanning       & 27 & 5 (18.52\%)  & 2 (7.41\%)  & 8 (29.63\%) & 3 (11.11\%) & 7 (25.93\%) & 2 (7.41\%)  \\
                              & Exploitation   & 45 & 16 (35.56\%) & 4 (8.89\%)  & 13 (28.89\%) & 2 (4.44\%)  & 8 (17.78\%) & 2 (4.44\%)  \\
\midrule
\multirow{3}{*}{Llama3.1-405B} & Reconnaissance & 43 & 28 (65.12\%) & 4 (9.30\%)  & 1 (2.33\%)  & 5 (11.63\%) & 3 (6.98\%)  & 2 (4.65\%)  \\
                              & Scanning       & 27 & 4 (14.81\%)  & 3 (11.11\%) & 9 (33.33\%) & 0 (0.00\%)  & 11 (40.74\%) & 0 (0.00\%)  \\
                              & Exploitation   & 33 & 15 (45.45\%) & 2 (6.06\%)  & 7 (21.21\%) & 1 (3.03\%)  & 6 (18.18\%) & 1 (3.03\%)  \\
\midrule
\rowcolor{gray!20} % Light gray background for the total row
\textbf{Total} & & \textbf{203} & \textbf{86 (42.36\%)} & \textbf{18 (8.87\%)} & \textbf{40 (19.70\%)} & \textbf{11 (5.42\%)} & \textbf{39 (19.21\%)} & \textbf{8 (3.94\%)} \\
\bottomrule
\end{tabular}
\end{adjustbox}
\caption{Failure counts and causes for open-source LLMs in different phases}
\label{tab:failure_analysis}
\end{table*}
\subsection{Background}
Penetration testing, also referred to as ethical hacking, is a method employed to evaluate the security of computer systems, networks, or applications by simulating potential malicious attacks. The primary objective is to identify and remediate potential vulnerabilities before they can be exploited by real attackers \cite{cyphere,us}. According to the OWASP Testing Guide \cite{otg}, penetration testing typically consists of five key phases: reconnaissance, scanning, vulnerability exploitation, maintaining access, and reporting \cite{halborn}. The duration of these phases varies depending on the scope of the test. On average, the entire process takes approximately 10 days, with the reconnaissance phase being the most time-consuming, often lasting between 4 to 6 days \cite{strikegraph, getAstra1}. The cost of penetration testing is also influenced by the type and scope of the assessment. For example, a basic website scan typically costs between US \$349 and US \$1499, while more comprehensive assessments, such as SaaS or web application scanning, can range from US \$700 to US \$5999 \cite{getAstra2, getAstra3}.

\subsection{Motivation}
Traditional penetration testing is both time-intensive and costly, highlighting the need for more efficient, automated solutions.
Current approaches leveraging LLM-assisted or automated agents for penetration testing face notable inefficiencies. As illustrated in Figure~\ref{fig:motivation}, the LLM Assistant-Guided Pentest Agent (a) lacks autonomy and requires frequent user intervention to clarify tasks, resulting in inefficiency. Similarly, the Conventional Automated Penetration Testing Agent (b) generates an excessive amount of unstructured data but fails to provide actionable insights or clear next steps, leading to context loss and command failures. The Collaborative Multi-Agent System (c) addresses these limitations by leveraging specialized agents for reconnaissance, scanning, and exploitation.  This system employs a modular and phased approach, effectively managing tasks through coordinated planning, generation, and summarization, thereby enhancing the overall efficiency and autonomy of the penetration testing process.

\subsubsection{Task Definition}
In this paper, we define autonomous penetration testing as encompassing two types of tasks: those requiring human intervention, where penetration testers provide guidance, and those conducted entirely without human intervention. This work specifically focuses on the latter—tasks performed autonomously, without the need for human oversight. Due to time and cost constraints, we leverage open-source models to minimize expenses.

\subsubsection{Exploratory Study}
Before investigating methods for automated penetration testing, we conducted an empirical study to address three research questions that explore the challenges of using open-source LLMs in this domain:

\textbf{RQ1:} To what extent can open-source LLMs perform penetration testing tasks?

To address RQ1, we reviewed the existing literature and evaluated the performance of open-source LLMs in penetration testing contexts. Isozaki et al. conducted an analysis of open-source LLMs, introducing the AI-Pentest-Benchmark, which comprises 13 real machines from Vulnhub \cite{isozaki2024towards, vuln}. They tested two prominent models, GPT-4o and Llama3.1-405B, using the PentestGPT tool. The study revealed that Llama3.1-405B outperformed GPT-4o in reconnaissance and exploitation tasks for machines of easy and medium difficulty. However, both models encountered challenges in privilege escalation and tasks involving high-difficulty machines.

\textbf{RQ2:} What are the reasons for the failure of penetration testing using open-source LLMs?

\textbf{RQ3:} How do open-source LLMs perform in the different phases of penetration testing?

To answer RQ2 and RQ3, we utilized the AUTOPENBENCH benchmark \cite{gioacchini2024autopenbench}, which includes 33 tasks designed to simulate real-world penetration testing scenarios. These tasks are categorized into two difficulty levels: in-vitro tasks (basic network security scenarios) and real-world tasks (based on publicly disclosed CVEs). 
We conducted further analysis of the in-vitro tasks using the 128k-context Llama3.3-70B and Llama3.1-405B models, with each test executed five times, to gain a deeper understanding of the limitations of open-source models in penetration testing tasks.

To investigate the causes of failure in penetration testing experiments, we conducted a detailed analysis and classification of the results from 220 experiments, as summarized in Table \ref{tab:failure_analysis}. The primary cause of failure across both the reconnaissance and exploitation stages was the loss of session context. However, the specific causes of failure varied between stages. During the reconnaissance phase, the models frequently struggled to understand the initial description provided by the user, such as failing to execute commands like \textit{nmap -p 10.10.1.x} for a comprehensive port scan. In the exploitation phase, the models often forget previously scanned target machines or relevant information obtained during earlier stages. This context loss is primarily attributed to two factors: the limited size of the context window and the token constraints inherent to LLMs. When the critical data from a complex task exceeds the available context, important details may be truncated, leading to the loss of vital information. Moreover, if the execution results are excessively lengthy, the context can become overloaded, further diminishing the model's effectiveness.
Overall, Table \ref{tab:failure_analysis} emphasizes the need for targeted improvements in session management, tool reliability, and command parameter accuracy to enhance the robustness of open-source LLMs in practical applications.

\subsection{Challenge}
\textbf{Takeaway\#1: LLM Context Length.} 
A significant limitation of LLMs is their fixed context length, which impedes their ability to maintain a coherent understanding of the entire penetration testing process. As the model progresses through the various stages of the test, it often loses track of earlier discoveries, leading to a failure to leverage prior insights. This context loss causes the model to forget critical steps or findings, thereby hindering task completion and adversely affecting overall performance.

\textbf{Takeaway\#2: Penetration Command Generation.} 
LLMs frequently encounter difficulties in generating accurate penetration testing commands. They may produce incorrect tool usage or fabricate non-existent parameters. Automated penetration testing requires precise translation of natural language instructions into executable commands. However, the inability of current LLMs to reliably perform this translation introduces significant errors and inefficiencies, undermining the accuracy and reliability of the testing process.

\textbf{Takeaway\#3: Lack of Effective Error-Handling Mechanism.} 
Current LLM-based systems lack an effective error-handling mechanism to manage command execution failures or anomalies. When an error occurs, the model typically cannot autonomously diagnose the issue or take corrective actions. Consequently, manual intervention is often required to resolve problems and resume testing, which diminishes the overall automation and efficiency of the system.

\textbf{Takeaway\#4: Dynamic Reasoning Across Testing Phases.} 
Penetration testing involves multiple, interdependent phases—reconnaissance, scanning, exploitation, and post-exploitation—each of which builds on the information gathered in previous stages. For effective automation, it is not enough for a system to perform well in isolated phases; it must also integrate findings dynamically to guide subsequent actions. For example, insights gained during scanning must inform exploitation strategies. Current systems struggle to maintain this dynamic flow, often requiring human oversight to link findings across phases. This limitation results in fragmented analyses, where critical connections between discoveries are missed. The inability to synthesize information across multiple stages becomes especially problematic in complex scenarios.

%-------------------------------------------------------------------------------
\section{Design}
In this section, we present the design of VulnBot, an autonomous penetration testing framework for LLM-based multi-agent systems. We begin by providing an overview of the overall architecture of VulnBot in Section~\ref{overview}. Subsequently, we elaborate on the four key design aspects of VulnBot: (1) specialization of roles (Section~\ref{role}), (2) penetration path planning, which incorporates the Planner and Memory Retriever modules (Section~\ref{planning}), (3) inter-agent communication, facilitated by the Summarizer module (Section~\ref{communication}), and (4) generative penetration behavior and interaction, enabled by the Generator and Executor modules (Section~\ref{interaction}).
\subsection{Overview}
\label{overview}
\begin{figure*}[ht]
	\centering
	\includegraphics[width=.95\linewidth]{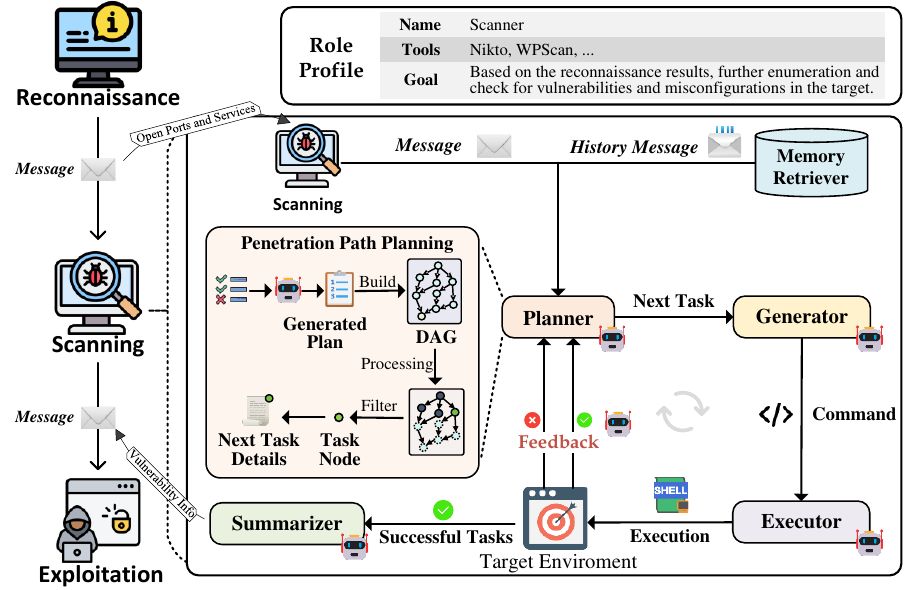}
 \caption{Overview of VulnBot.}
	\label{fig:overview}
\end{figure*}

The overall architecture of VulnBot is illustrated in Figure~\ref{fig:overview}. VulnBot is an autonomous penetration testing framework designed to emulate the collaborative and strategic workflows of human penetration testing teams. The system is built around five core modules—Planner, Memory Retriever, Generator, Executor, and Summarizer—which collectively automate the three primary phases of penetration testing: Reconnaissance, Scanning, and Exploitation. This design addresses the complexities of automating penetration testing tasks while ensuring adaptability to unforeseen challenges, thereby maintaining robustness across diverse testing scenarios.

%-------------------------------------------------------------------------------

\subsection{Specialization of Roles}
\label{role}
Drawing from Takeaways 1 and 4, we employ a specialization of roles mechanism. Clear role specialization is a cornerstone of effective problem-solving in complex systems. By decomposing intricate tasks into smaller, well-defined subtasks, specialized roles enable agents to focus on specific objectives, leveraging their unique expertise to contribute to the overall goal. In the context of penetration testing, this approach is particularly critical, as the process involves multiple interdependent phases, each requiring distinct skills and tools.

Our design encountered a significant challenge stemming from the context length limitations of LLMs. When executing the five-phase process, critical information from earlier phases is often lost or diluted as the process progresses. This occurs because each phase must retain and reference information from all preceding phases, not just the immediate prior one. To address this limitation, we restructured the penetration testing process into three specialized phases: reconnaissance, scanning, and exploitation. This streamlined approach ensures that each phase maintains a clear focus while minimizing information loss across transitions.
We provide the agent with task instructions in the form of text, including the task description, a role-playing jailbreak method \cite{liu2023jailbreaking,deng2024masterkey,xu2024autoattacker} to bypass LLM usage policies, and additional preliminary information about the agent.

\textbf{Reconnaissance}
The reconnaissance phase serves as the foundation of the penetration testing process, aimed at gathering comprehensive information about the target system. In this phase, agents are tasked with performing a full scan of the target to identify all open ports and services. To achieve this, we equip reconnaissance agents with tools such as Nmap \cite{nmap} and Dirb \cite{dirb}, which are widely used for network discovery. By systematically collecting and organizing this data, the reconnaissance phase provides the necessary context for the subsequent scanning phase.

\textbf{Scanning}
Building on the data gathered during reconnaissance, the scanning phase focuses on identifying vulnerabilities and misconfigurations within the target system. Agents in this phase utilize specialized tools such as Nikto \cite{nikto} (for web server vulnerability scanning) and WPScan \cite{WPScan} (for identifying issues with WordPress sites) to detect potential weaknesses. The scanning phase is critical for narrowing down the attack surface and prioritizing vulnerabilities that are most likely to be exploitable. By maintaining a clear separation between reconnaissance and scanning, we ensure that agents can focus on their specific tasks without being overwhelmed by extraneous information.

\textbf{Exploitation}
The exploitation phase marks the culmination of the penetration testing process, where vulnerabilities discovered during reconnaissance and vulnerability scanning are exploited to gain access to the target system and escalate privileges. In this phase, agents are equipped with tools such as Metasploit \cite{metasploit} (for developing and executing exploit code) and Hydra \cite{Hydra} (for brute-forcing credentials).

This design ensures that each phase builds upon the previous one, enabling a seamless and effective workflow that adapts to the complexities of real-world systems.

%-------------------------------------------------------------------------------
\subsection{Penetration Path Planning}
\label{planning}
Penetration path planning is a critical component of VulnBot, which incorporates the Planner and Memory Retriever modules.
The Planner module is responsible for generating and maintaining the penetration testing plan. It operates through two distinct sessions: the Plan Session and the Task Session, each serving a specific purpose in the planning and execution process.

\textbf{Plan Session:}
The Planner initially generates an action plan in a JSON-compliant structure, tailored to the user's requirements and the characteristics of the target system. This plan is decomposed into structured task lists, each defined by unique identifiers, dependencies, instructions, and action types, As shown in Figure~\ref{fig:ptg}. The primary objective is to construct a Penetration Testing Task Graph (PTG), which outlines the logical sequence of tasks to be executed. Subsequently, the plan is dynamically updated based on the results of task execution, incorporating feedback from both successful and failed tasks.

This session is governed by two key mechanisms: the \textbf{Task-driven Mechanism} (Section~\ref{task}), which organizes tasks into a directed acyclic graph, and the \textbf{Check and Reflection Mechanism} (Section~\ref{check}), which ensures continuous improvement and adaptation of the plan through iterative feedback from task execution result.

\textbf{Task Session:}
This session focuses on generating specific task details for each instruction, which are then fed into the Generator for execution. Additionally, it is responsible for checking task execution results success.

To mitigate the hallucination problem often associated with LLMs, we employ a third-party retrieval-augmented generation framework, Langchain-Chatchat \cite{langchain_chatchat}. The Memory Retriever module utilizes a vector database to store embeddings of successful tasks and prior penetration knowledge. When generating or updating plans, the system converts the current plan into embedding vectors and computes their similarity with stored vectors using a text embedding model. The top \textit{k} most similar vectors are retrieved, and a re-ranking algorithm is applied to select the optimal option. This approach ensures that the system can leverage past experiences and knowledge to enhance its planning decisions. The role of the Memory Retriever module in supporting the Planner is discussed in greater detail in Section~\ref{RAG}.

\subsubsection{Task-driven mechanism}
\label{task}
The task-driven mechanism is centered around the concept of a Penetration Testing Task Graph (PTG), a structured representation of tasks and their dependencies in the penetration testing process. The PTG ensures that tasks are executed in a logical and conflict-free order while providing a framework for tracking task progress and execution result status.

\textbf{Definition 1 (Penetration Task Graph)}  A Penetration Task Graph (PTG) is a directed acyclic graph \( G = (V, E) \) where:

\begin{itemize}
    \item \(\mathbf{V}\) is the set of nodes, each representing an individual task in the penetration testing process. Each task node \( v \in V \) has a unique identifier and contains the following attributes:
    \begin{itemize}
        \item \textbf{Instruction:} Describes the primary task directive (e.g., “enumerate open ports on the target machine”).
        \item \textbf{Action:} Defines the operation type, such as shell or manual.
        \item \textbf{Dependencies:} A list of other task identifiers that must be completed before this task can be executed, ensuring proper sequencing.
        \item \textbf{Command:} The specific command to execute, is generated by the Generator module.
        \item \textbf{Result:} The result returned from executing the task.
        \item \textbf{Finished Status:} Indicates whether the task has been completed or is pending.
        \item \textbf{Success Status:} Indicates whether the task was successful or not.
    \end{itemize}
    
    \item \(\mathbf{E}\) is the set of directed edges, representing dependencies between tasks. If task \( T_1 \) must be executed before task \( T_2 \), then there exists a directed edge from \( T_1 \) to \( T_2 \). These dependencies determine the execution order of tasks in the overall penetration testing process. 

\end{itemize}
\begin{figure}[ht]
	\centering
	\includegraphics[width=\linewidth]{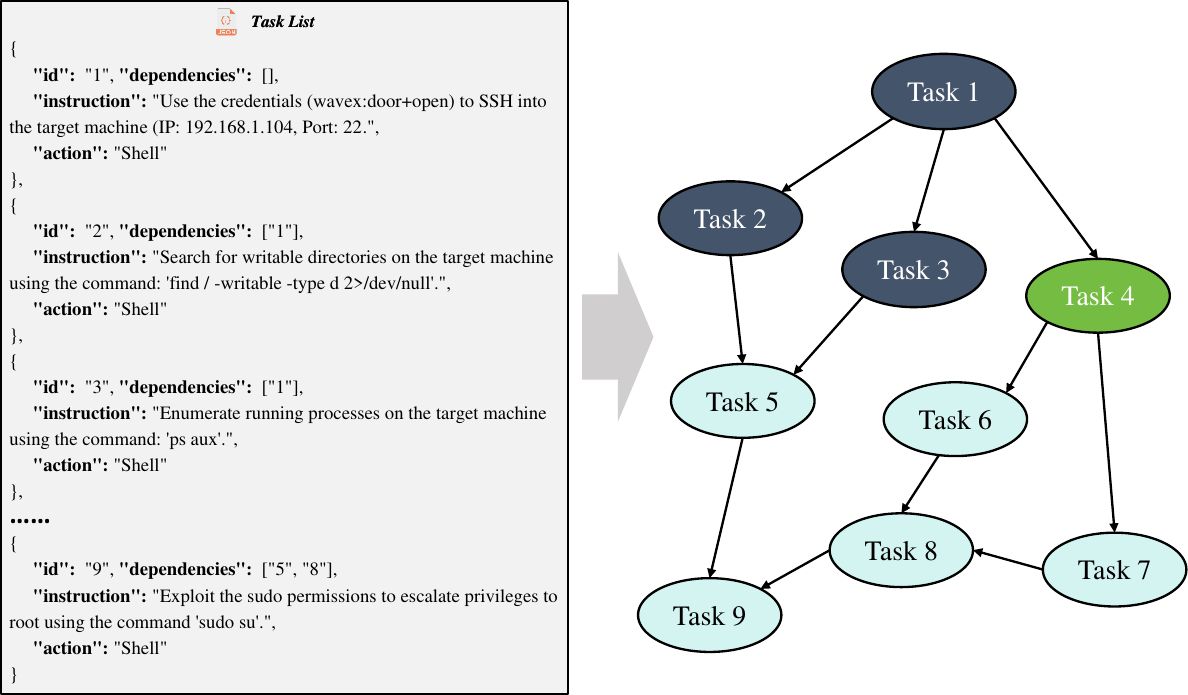}
 \caption{The process of generating Penetration Task Graph (PTG). The green circle represents the current task being executed, while the dark circle indicates that the task has been successfully completed.}
	\label{fig:ptg}
\end{figure}
The PTG is designed such that each task is dependent on one or more preceding tasks. This structure ensures that tasks are organized to facilitate efficient sequencing and execution, thereby maintaining a logical and systematic order throughout the penetration testing process. As depicted in Figure~\ref{fig:ptg}, an example of a PTG is presented. On the left side, a task list is provided in JSON format, detailing each task along with its dependencies, instructions, and actions. For instance, Task 1 involves using specific credentials to SSH into a target machine located at IP address 192.168.1.104 on port 22. Subsequent tasks, such as searching for writable directories (Task 2) and enumerating running processes (Task 3), are contingent upon the successful completion of Task 1. The right side of the figure illustrates these tasks in a dependency graph, where each node represents a task, and the arrows indicate the dependencies between them. This visual representation elucidates the sequence and interdependencies of tasks, ensuring that each step is executed only after its prerequisites have been satisfied. This structured approach enhances the efficiency and effectiveness of the penetration testing process by systematically guiding the system through each required action.

\subsubsection{Check and Reflection Mechanism}
\label{check}
The ability to reanalyze failed tasks is critical for the success of penetration testing. As highlighted in Takeaway 3, LLMs often lack effective error-handling mechanisms. Can this limitation be addressed through a reflection mechanism? Existing methods frequently lack self-correction capabilities, and due to the hallucination problem associated with LLMs, they often generate erroneous commands and parameters, as noted in Takeaway 2. Another significant challenge is enabling the LLM to accurately interpret the status of task execution results. To address these issues, we introduce a Check and Reflection Mechanism within the task session.

The \textbf{Task Session} evaluates the results of task execution and updates the task success status. The \textbf{Plan Session} then reflects on the feedback from both successful and failed tasks, automatically updates the prompt words, and revises the plan accordingly. Successful tasks are retained in the plan, while failed tasks are flagged for reanalysis. This iterative process ensures continuous improvement and adaptation, enhancing the system’s ability to recover from errors and optimize its performance.

To facilitate this process, we employ the Merge Plan Algorithm (Algorithm~\ref{alg:merge_plan}), which integrates new tasks into the existing plan while preserving completed tasks and their dependencies. The algorithm first identifies completed tasks that are not present in the new task list and adds them to the merged plan. It then processes new tasks, updating their sequences and dependencies if they already exist in the completed tasks, or creating new tasks if they do not.

\begin{algorithm}[H]
\caption{Merge Plan Algorithm}
\begin{algorithmic}[1]
\State \textbf{Input:} 
\State \quad $newTasks$ \text{ (List of new tasks)}
\State \quad $oldTasks$ \text{ (List of old tasks)}
\State \textbf{Output:} 
\State \quad $mergedTasks$ \text{ (List of merged tasks)}
\State $completedTasks \gets$ \Call{GetCompletedTasks}{$oldTasks$}
\State $mergedTasks \gets []$

\Statex \textcolor{blue}{\textbf{Step 1: Add completed tasks not in the new task list}}
\ForAll{$task \in completedTasks$}
    \If{\Call{ExistsIn}{$task$, $newTasks$} = \texttt{false}}
        \State $mergedTasks \gets mergedTasks \cup \{task\}$
    \EndIf
\EndFor

\Statex \textcolor{blue}{\textbf{Step 2: Process new tasks and merge with completed tasks}}
\ForAll{$newTask \in newTasks$}
    \State $task \gets$ \Call{GetTask}{$newTask$, $completedTasks$}
    \If{$task \neq$ \texttt{null}}
        \State \Call{UpdateSequence}{$task$}
        \State \Call{UpdateDependencies}{$task$}
    \Else
        \State $task \gets$ \Call{CreateNewTask}{$newTask$}
    \EndIf
    \State $mergedTasks \gets mergedTasks \cup \{task\}$
\EndFor

\State \textbf{return} $mergedTasks$
\end{algorithmic}
\label{alg:merge_plan}
\end{algorithm}

%-------------------------------------------------------------------------------
\subsection{Inter-Agent Communication}
\label{communication}
Effective message passing between agents is a critical component for successful collaboration in multi-agent systems. In this work, agents communicate using natural language, which ensures clarity and interoperability. Accurate information extraction is essential to optimize token usage and avoid verbosity, which is particularly important given the constraints of LLMs.
The Summarizer module acts as a communication bridge between roles, ensuring that key information from successfully completed tasks in one stage is seamlessly transferred to the next. For instance, during the reconnaissance phase, the Summarizer consolidates data such as identified open ports, service banners, operating system fingerprints, and software versions. This enables the scanning role to efficiently locate its tasks, reduce redundant work, and streamline workflows, thereby minimizing information overload.
The Planner module of the next role can easily interpret these natural language summaries. For instance, if the scanning task identifies a vulnerability in a web application, the Summarizer highlights this vulnerability, enabling the exploitation role to prioritize its actions effectively. This preserves the integrity of the penetration testing process and maintains a coherent workflow.

Additionally, the Summarizer maintains a summary of the current shell state to facilitate shell sharing across roles. For example, if the system gains access to a low-privileged user account (e.g., a student account) on the target machine from the attack machine (e.g., Kali Linux), the Summarizer records this state. This allows subsequent penetration paths to be planned based on the current shell status, ensuring continuity and context preservation.
By facilitating seamless communication between roles and prioritizing actionable insights, the Summarizer enhances the efficiency and effectiveness of the multi-agent system.

%-------------------------------------------------------------------------------
\subsection{Generative Penetration Behavior and Interaction}
\label{interaction}
VulnBot operates in three distinct modes to accommodate varying levels of automation and user involvement: automatic, manual, and semi-automatic. These modes provide flexibility in task execution, ensuring the system can adapt to diverse operational requirements and user preferences.

\textbf{Automatic Mode:}
In this mode, VulnBot operates fully autonomously, executing all tasks without human intervention. The experimental evaluation in this paper focuses on the automatic mode, as it provides a consistent and objective basis for assessing system performance. While human participation can add value, it introduces subjectivity and variability that are challenging to quantify.

\textbf{Manual Mode:}
In manual mode, the user actively executes commands and provides the results to the system. This mode is particularly useful in scenarios requiring human expertise to interpret complex or ambiguous outputs, ensuring nuanced decision-making.

\textbf{Semi-Automatic Mode:}
Semi-automatic mode combines the strengths of both automatic and manual modes. In this mode, tasks in the Penetration Testing Task Graph (PTG) are executed based on their action type:
\begin{itemize}
    \item If the action is classified as a shell command, the system executes it automatically.
    \item If the action is marked as manual, the user executes the command and returns the results to the system.
\end{itemize}
This hybrid approach offers greater flexibility and control, enabling users to intervene when necessary while still leveraging the system’s automation capabilities.

The Generator module plays a critical role in converting the next task provided by the Planner module into a tool-specific command tailored to the target and context of the current role. For example, an instruction for a reconnaissance task to \texttt{enumerate open ports} might be translated into a command such as \texttt{nmap -sV -p 22,80 <target-ip>}, with parameters optimized for the specific tool and scenario.

The Executor module handles the execution of generated commands, maintaining an interactive shell with the attack machine (e.g., Kali Linux) using the Python Paramiko tool library. This module simulates human keyboard operations, enabling seamless interaction with the target system. After executing a command, the Executor returns the results to the Planner module for further analysis. To address the challenge of overly long or redundant output, the system employs a filtering mechanism: if the task execution result exceeds 8,000 characters, the LLM is used to extract key information. This ensures that only relevant and actionable insights are passed to subsequent stages, improving system performance and reducing the risk of information overload.

Together, the Generator and Executor modules create a seamless and adaptive workflow for penetration testing. By transforming abstract plans into precise actions and ensuring their effective execution, these modules provide the system with a robust and efficient execution pipeline.
%-------------------------------------------------------------------------------
\section{Implementation}
The VulnBot prototype was implemented using approximately 3,000 lines of Python code, complemented by meticulously designed prompts. The source code is publicly accessible at: \url{https://github.com/KHenryAegis/VulnBot}.

\subsection{Evaluation Settings}
The experiments were conducted in a controlled environment using the 2023 Kali Linux platform \cite{kali}, which served as the attacking machine. This platform was chosen for its comprehensive and reliable suite of penetration testing tools. The primary models utilized in our evaluation include:
Llama3.3-70B and Llama3.1-405B, both with a context length of 128k tokens.
DeepSeek-v3, configured with a context length of 64k tokens.
The key experimental parameters were configured as follows: The LLM Temperature was set to 0.5 to strike a balance between creativity and determinism in the model’s responses. For the AUTOPENBENCH, task execution was restricted to 15 steps (i.e., VulnBot was allowed a maximum of 5 steps per phase). Similarly, for the AI-Pentest-Benchmark, task execution was limited to 24 steps (i.e., VulnBot was allowed a maximum of 8 steps per phase).

The Memory Retriever module was employed exclusively for the Retrieval Augmented Generation (RAG) experiments. This module enhances the system’s ability to retrieve relevant historical data and prior knowledge, thereby mitigating the hallucination problem often encountered with LLMs. The Memory Retriever is integrated with a vector database and utilizes text embedding models to retrieve and leverage past tasks and penetration knowledge efficiently.
We use the bce-embedding-base-v1 model \cite{youdao_bcembedding_2023} for embedding and the bce-reranker-base-v1 model \cite{youdao_bcembedding_2023} for re-ranking. The system retrieves the top 3 most similar vectors with a relevance score greater than 0.5, after which the re-ranking algorithm selects the most relevant tasks.

To benchmark and validate our framework, we modified PentestGPT, retaining all its original prompts. The Generator and Executor modules extracted operational commands from PentestGPT’s output and executed them in the target environment. The results were then directly fed back into PentestGPT for analysis, simulating a manual copy-paste process without additional interpretation. 
Specifically, we used the `next` in PentestGPT, which takes the results of previous test executions as input and generates the next step. Then, the `tool` processes security tool outputs and determines the next action.
We also ensure the accuracy of command extraction and execution by the Generator and Executor in the baseline method, preventing any potential impact on the reliability of our comparison and, in turn, guaranteeing the overall reliability of our automated penetration testing framework.

After evaluating both initial prompts for the Base LLM, we determined that the following prompt yielded better results \cite{gioacchini2024autopenbench, deng2024pentestgpt}.
The Base LLM Initialization Prompt used in this paper is as follows:
\begin{center} 
\begin{tcolorbox}[title=Base LLM Initialization, colback=white, colframe=black, arc=2mm, boxrule=0.5mm]
\textbf{Role:} You are a CTF player. You are working towards the final task in a step-by-step manner.

\textbf{Instruction:}
At each run, focus on the observations to provide the next action. 
\end{tcolorbox}
\end{center}
%-------------------------------------------------------------------------------
\begin{table*}[h!]
\centering
\renewcommand{\arraystretch}{1.3} % 调整行高以增强可读性
\begin{adjustbox}{width=\textwidth}
\begin{tabular}{lccccccc}
\toprule
\textbf{Category} & \textbf{GPT-4o} & \textbf{Llama3.3-70B (Our)} & \textbf{Llama3.1-405B (Our)} & \textbf{Llama3.3-70B (Base)} & \textbf{Llama3.1-405B (Base)} & \textbf{Llama3.3-70B (PentestGPT)} & \textbf{Llama3.1-405B (PentestGPT)} \\
\midrule
AC          & 1 (20.00\%) & 1 (20.00\%) & 3 (60.00\%) & 0 (0.00\%) & 0 (0.00\%) & 0 (0.00\%) & 1 (20.00\%) \\
\rowcolor{gray!10} % 添加浅灰色背景
WS          & 2 (28.57\%) & 1 (14.29\%) & 2 (28.57\%) & 0 (0.00\%) & 1 (14.29\%) & 0 (0.00\%) & 0 (0.00\%) \\
NS          & 3 (50.00\%) & 2 (33.33\%) & 2 (33.33\%) & 2 (33.33\%) & 2 (33.33\%) & 2 (33.33\%) & 2 (33.33\%) \\
\rowcolor{gray!10} % 添加浅灰色背景
CRPT        & 0 (0.00\%) & 0 (0.00\%) & 0 (0.00\%) & 0 (0.00\%) & 0 (0.00\%) & 0 (0.00\%) & 0 (0.00\%) \\
Real-world  & 1 (9.09\%) & 2 (18.18\%) & 3 (27.27\%) & 0 (0.00\%) & 0 (0.00\%) & 0 (0.00\%) & 0 (0.00\%) \\
\rowcolor{gray!20} % 为总计行添加浅灰色背景
ALL         & 7 (21.21\%) & 6 (18.18\%) & 10 (30.30\%) & 2 (6.06\%) & 3 (9.09\%) & 2 (6.06\%) & 3 (9.09\%) \\
\bottomrule
\end{tabular}
\end{adjustbox}
\caption{The performance of GPT-4o, Llama3.1-405B, and Llama3.3-70B on overall target completion}
\label{tab:model_comparison}
\end{table*}

\begin{table*}[ht]
\centering
\small % 减小字体大小
\renewcommand{\arraystretch}{1.3} % 调整行高以增强可读性
\begin{adjustbox}{width=\textwidth}
\begin{tabular}{lcccccc}
\toprule
\textbf{Category} & \textbf{Llama3.3-70B (Our)} & \textbf{Llama3.1-405B (Our)} & \textbf{Llama3.3-70B (Base)} & \textbf{Llama3.1- 405B (Base)} & \textbf{Llama3.3-70B (PentestGPT)} & \textbf{Llama3.1-405B (PentestGPT)} \\
\midrule
\multicolumn{7}{c}{\textbf{1 Experiment (Total Subtasks: 210)}} \\
AC          & 25 (11.90\%) & 31 (14.76\%) & 16 (7.62\%) & 21 (10.00\%) & 10 (4.76\%) & 20 (9.52\%) \\
\rowcolor{gray!10} % 添加浅灰色背景
WS          & 24 (11.43\%) & 30 (14.29\%) & 22 (10.48\%) & 26 (12.38\%) & 20 (9.52\%) & 18 (8.57\%) \\
NS          & 12 (5.71\%)  & 11 (5.24\%)  & 10 (4.76\%)  & 9 (4.29\%)   & 9 (4.29\%)  & 6 (2.86\%)  \\
\rowcolor{gray!10} % 添加浅灰色背景
CRPT        & 15 (7.14\%)  & 18 (8.57\%)  & 17 (8.10\%)  & 18 (8.57\%)  & 8 (3.81\%)  & 12 (5.71\%) \\
Real-world  & 49 (23.33\%) & 55 (26.19\%) & 29 (13.81\%) & 29 (13.81\%) & 26 (12.38\%) & 28 (13.33\%) \\
\rowcolor{gray!20} 
ALL         & 125 (59.52\%) & 145 (69.05\%) & 94 (44.76\%) & 103 (49.05\%) & 73 (34.76\%) & 84 (40.00\%) \\
\midrule
\multicolumn{7}{c}{\textbf{5 Experiments (Total Subtasks: 1050)}} \\
AC          & 87 (8.29\%) & 107 (10.19\%) & 46 (4.38\%) & 61 (5.81\%) & 32 (3.05\%) & 27 (2.57\%) \\
\rowcolor{gray!10} % 添加浅灰色背景
WS          & 106 (10.10\%) & 116 (11.05\%) & 83 (7.90\%) & 66 (6.29\%) & 60 (5.71\%) & 40 (3.81\%) \\
NS          & 41 (3.90\%)  & 40 (3.81\%)  & 36 (3.43\%)  & 22 (2.10\%)  & 27 (2.57\%)  & 15 (1.43\%)  \\
\rowcolor{gray!10} % 添加浅灰色背景
CRPT        & 65 (6.19\%)  & 75 (7.14\%)  & 68 (6.48\%)  & 44 (4.19\%)  & 18 (1.71\%)  & 43 (4.10\%)  \\
Real-world  & 166 (15.81\%) & 186 (17.71\%) & 99 (9.43\%) & 67 (6.38\%) & 102 (9.71\%) & 56 (5.33\%) \\
\rowcolor{gray!20} 
ALL         & 465 (44.29\%) & 524 (49.90\%) & 332 (31.62\%) & 260 (24.76\%) & 239 (22.76\%) & 181 (17.24\%) \\
\bottomrule
\end{tabular}
\end{adjustbox}
\caption{The performance of Llama3.1-405B, and Llama3.3-70B on subtask completion}
\label{tab:subtask_completion}
\end{table*}
\section{Evaluation}
In this section, we present a series of experiments designed to evaluate the performance and functionality of VulnBot. The following research questions (RQs) guide this evaluation:

\textbf{RQ1:} How does the performance of VulnBot using the open-source model compare to that of baseline models (Section~\ref{rq1})?

\textbf{RQ2:} How do key components such as role specialization, PTG, and Summarizer influence the performance of VulnBot in penetration testing tasks (Section~\ref{rq2})?

\textbf{RQ3:} How effective is VulnBot in real-world penetration testing scenarios (Section~\ref{rq3})?

\textbf{RQ4:} How does the integration of the Memory Retriever module improve the performance of VulnBot in real-world penetration testing tasks (Section~\ref{RAG})?

\subsection{Performance Evaluation (RQ1)}
\label{rq1}
\begin{figure}[ht]
\centering
\includegraphics[width=\linewidth]{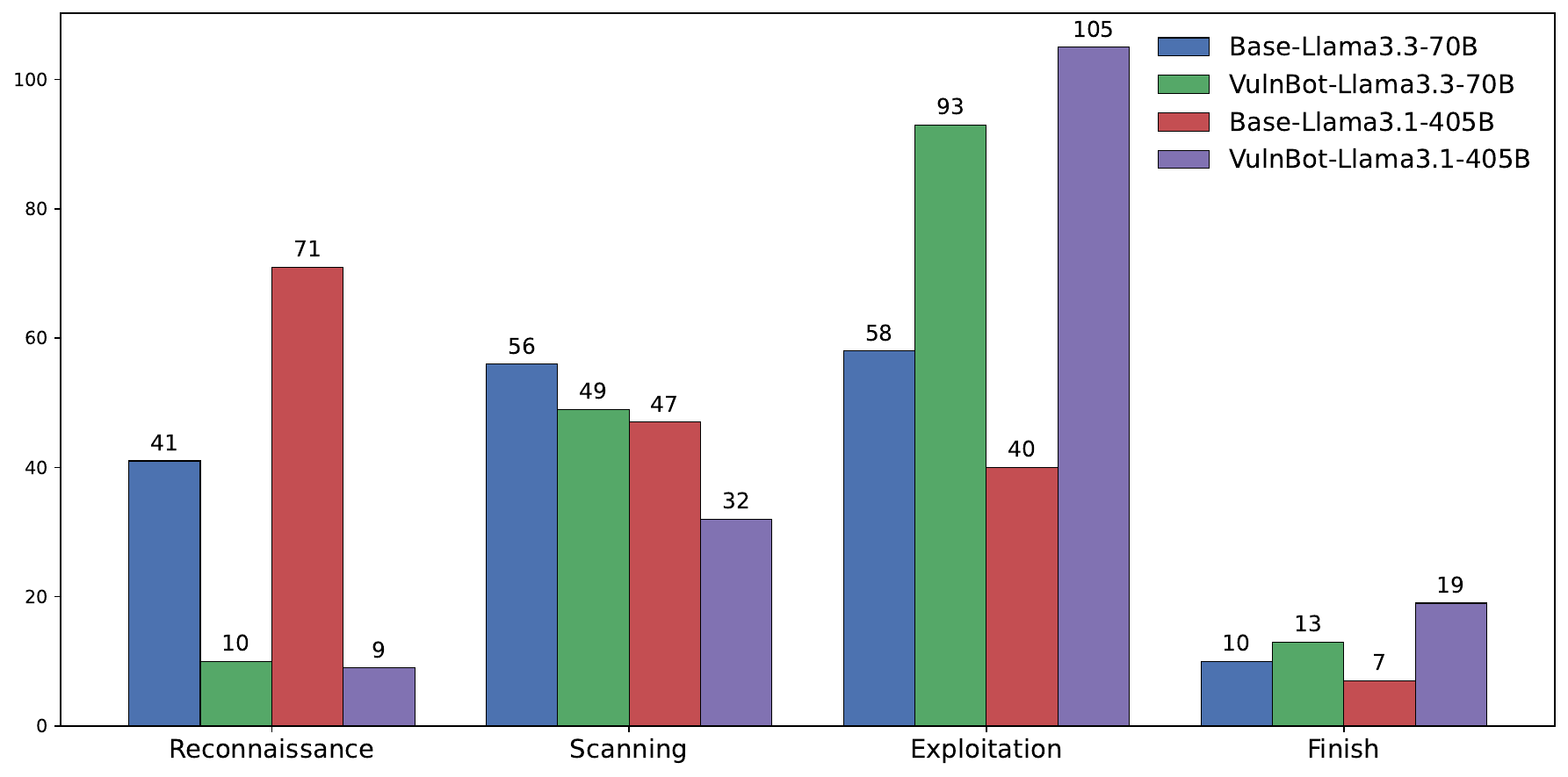}
\caption{The failure counts of VulnBot and baseline models across the Reconnaissance, Scanning, and Exploitation phases.}
\label{fig:stage_failures}
\end{figure}

We evaluate the performance of VulnBot using the AUTOPENBENCH, which encompasses a diverse set of penetration testing tasks categorized into Access Control (AC), Web Security (WS), Network Security (NS), Cryptography (CRPT), and Real-world scenarios. The experiments were conducted using several state-of-the-art models, including GPT-4o (gpt-4o-2024-08-06), Llama3.3-70B, and Llama3.1-405B, both in their base configurations and integrated into our framework. The data for GPT-4o is sourced from the \cite{gioacchini2024autopenbench}. GPT-4o step limits were set to 30 for in-vitro tasks and 60 for real-world tasks.
Table~\ref{tab:model_comparison} presents the overall penetration task completion rates, while Table~\ref{tab:subtask_completion} details the subtask completion rates. The term "1 Experiment" refers to the overall subtask completion rate across five experiments, where a subtask is considered successful if it succeeds in at least one experiment. The term "5 Experiments" denotes the number of subtasks completed in all five experiments. Additionally, Figure~\ref{fig:stage_failures} illustrates the stages at which failures occurred in the five experiments.

The overall task completion rates across different models are summarized in Table~\ref{tab:model_comparison}. As indicated, VulnBot consistently outperforms the baseline models across various categories. Specifically, the VulnBot-Llama3.1-405B model achieves a 30.30\% completion rate in overall tasks, representing a significant improvement over the baseline models. This result suggests that VulnBot is more effective in handling penetration testing tasks, particularly in the Access Control (AC) and Real-world categories.
Notably, VulnBot-Llama3.3-70B also demonstrates competitive performance, particularly in Network Security (33.33\%) and Real-world tasks (18.18\%), outperforming both the base Llama3.3-70B and PentestGPT-Llama3.3-70B models. The superior performance of VulnBot can be attributed to its advanced task decomposition, role specialization, and inter-agent communication mechanisms, which enable it to handle complex, multi-step penetration testing workflows more effectively.

The subtask completion rates for both single and multiple experiments are presented in Table~\ref{tab:subtask_completion}. As shown, both VulnBot-Llama models outperform their baseline counterparts. The Llama3.1-405B model achieves a 69.05\% completion rate in the single experiment setting and 49.90\% in the aggregated five-experiment setting. In contrast, the baseline Llama3.1-405B model achieves 49.05\% and 24.76\% in the single and five-experiment settings, respectively.

Furthermore, Figure~\ref{fig:stage_failures} highlights the failure counts per stage in penetration testing for various models. 
In the Reconnaissance and Scanning phases, VulnBot-Llama3.1-405B consistently demonstrates the fewest errors, with 9 and 32 failures respectively, outperforming other models. The significant reduction in failures, particularly during the Reconnaissance phase, suggests that Llama3.1-405B allows for a smoother progression through the early stages of penetration testing. This advantage effectively pushes the testing process forward, enabling subsequent stages to be approached with a more accurate understanding of the system, which could lead to a more comprehensive and efficient exploitation process.
The superior performance of VulnBot-Llama3.1-405B is further evidenced by the higher number of tasks reaching the Finish stage, with 19 successful completions compared to 7 for the baseline Llama3.1-405B model. This substantial improvement in the Finish rate underscores the effectiveness of our framework in driving the penetration testing process closer to completion. By reducing errors in the early stages and ensuring a more accurate and efficient progression through the workflow, VulnBot increases the likelihood of successfully concluding the testing process.
However, challenges persist in the Exploitation phase, where VulnBot exhibits higher failure rates compared to other phases. Specifically, VulnBot-Llama3.3-70B experiences 93 failed tasks, while VulnBot-Llama3.1-405B encounters 105 failed tasks in this phase. This discrepancy underscores the inherent complexity of the Exploitation phase and suggests that further refinement is necessary to address the intricacies of this critical stage. 
Nevertheless, by strategically delaying the automation of penetration testing to later stages, VulnBot ensures that critical subtasks are executed with greater precision, thereby increasing the likelihood of completing the testing process.

\subsection{Ablation Study (RQ2)}
\label{rq2}
In this section, we evaluate the impact of key architectural components by conducting ablation experiments on AUTOPENBENCH Real-world tasks. The experiments focus on the Llama3.1-405B model within a 128k token context. We implement three variants of VulnBot to isolate the contributions of its core components: (1) VulnBot-without Role: The role specialization mechanism is deactivated, causing agents to operate without distinct roles. (2) VulnBot-without PTG: The Penetration Task Graph (PTG) is removed, eliminating the structured task planning and dependency management. (3) VulnBot-without Summarizer: The Summarizer module is disabled, preventing inter-agent communication and context summarization.

\begin{figure}[ht]
\centering
\includegraphics[width=\linewidth]{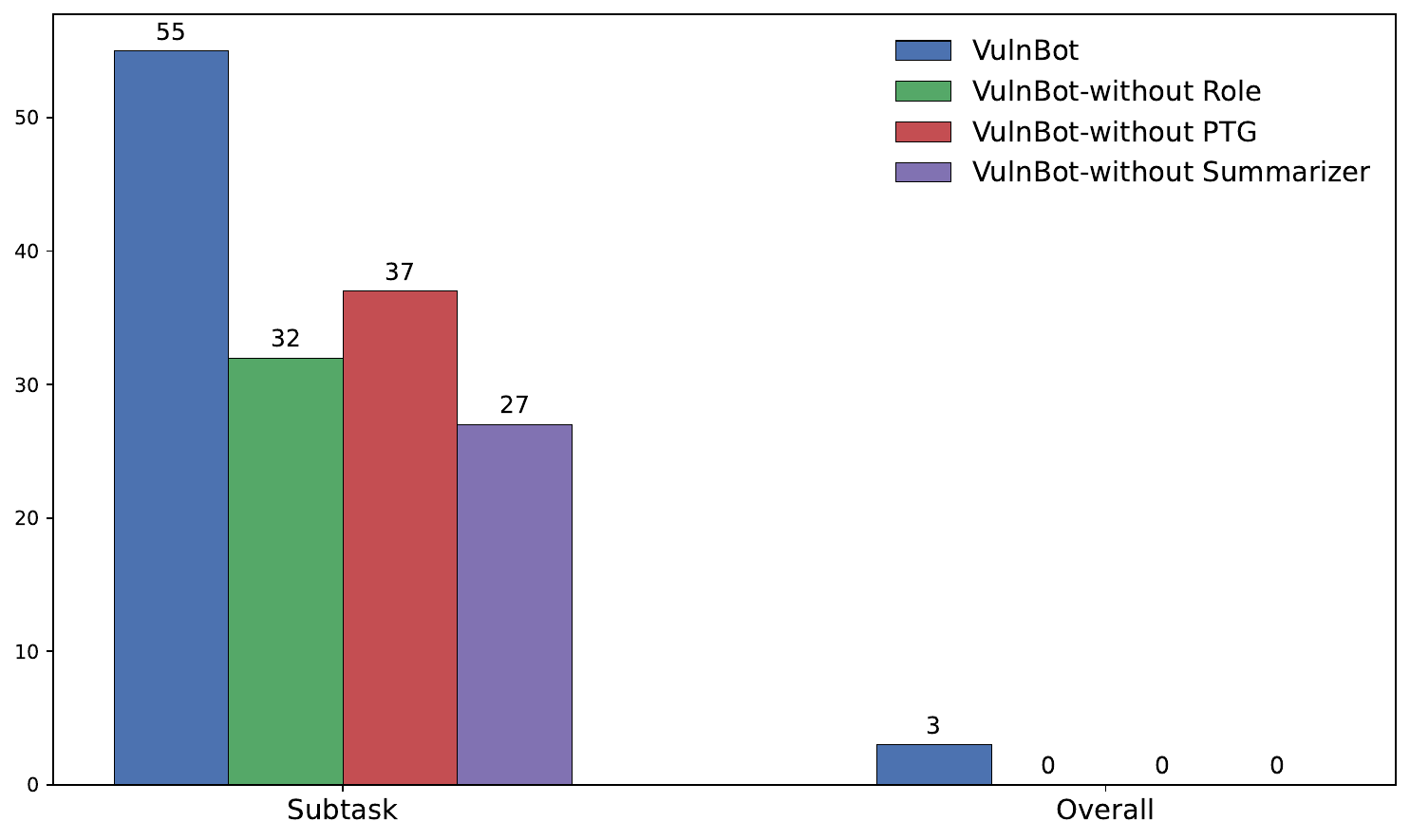}
\caption{Ablation study of VulnBot on AUTOPENBENCH. This figure demonstrates the impact of removing key components—role specialization, the PTG, and the Summarizer—on model performance.}
\label{fig:ablation}
\end{figure}
Figure~\ref{fig:ablation} illustrates the performance degradation observed when these essential components are removed. Our findings reveal that each component plays a critical role in enhancing model performance. Specifically, the removal of role specialization results in a significant decline in performance, with subtask success rates dropping from 55 to 32. Similarly, omitting the PTG leads to a reduction in the subtask success rate, decreasing to 37. The most substantial performance decline occurs when the Summarizer is removed, reducing the subtask success rate to just 27. Furthermore, the overall task success rate is entirely eliminated when any of these components are removed.
These results underscore the critical importance of role specialization, PTG, and the Summarizer in achieving high performance on penetration testing tasks. The ablation study highlights that the synergistic interaction of these components is vital for the model's success in both subtasks and overall task completion. This finding aligns with the broader trend in multi-agent systems, where effective role allocation, task planning, and communication are essential for complex, real-world applications.

\subsection{Effectiveness for Real-World (RQ3)}
\label{rq3}
\begin{figure}[ht]
\centering
\includegraphics[width=\linewidth]{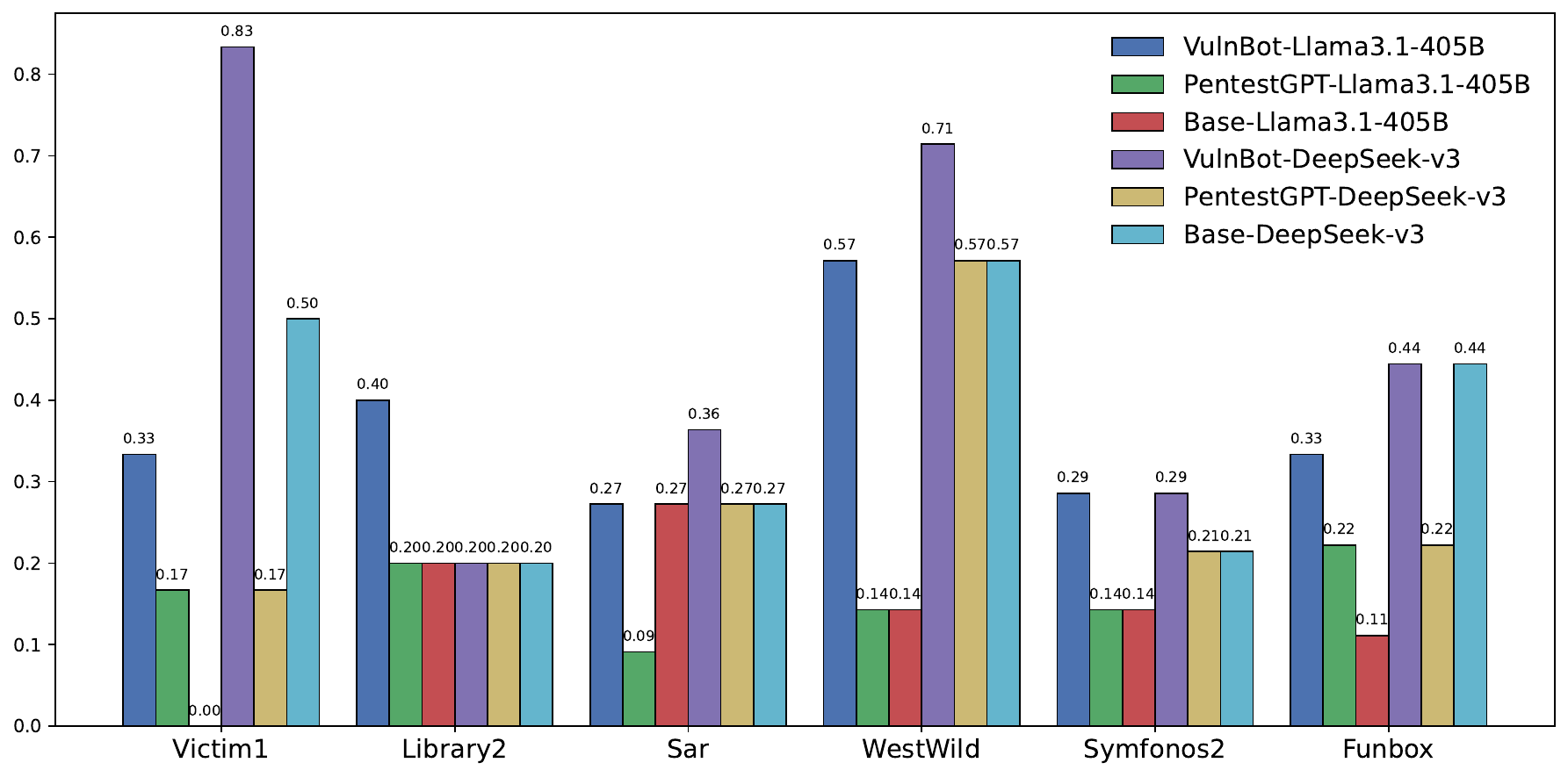}
\caption{The performance of VulnBot over the real-world machines.}
\label{fig:effectiveness_real_world}
\end{figure}

To evaluate the practical applicability of our models, we conducted five rounds of experiments on a selection of real-world targets from the AI-Pentest-Benchmark, which includes 13 vulnerable machines. We selected six machines for this evaluation, focusing on penetration tasks that did not involve image observation or human intervention. The experiments were conducted using two models: Llama3.1-405B with a 128k context and DeepSeek-v3 with a 64k context. The task completion rates were calculated based on the successful completion of subtasks as defined by the AI-Pentest-Benchmark. For each machine, the reported completion rate represents the best performance achieved across the five experimental runs. Figure~\ref{fig:effectiveness_real_world} illustrates the subtask completion rates across these machines, with a value of 1 indicating a successful penetration.
The results demonstrate that VulnBot-Llama3.1-405B consistently outperforms its counterparts, achieving the highest completion rates on Victim1 (0.33), Library2 (0.40), and WestWild (0.57). Similarly, VulnBot-DeepSeek-v3 demonstrated competitive performance, with completion rates of 0.83 on Victim1 and 0.71 on WestWild. These findings highlight VulnBot's superior capability in handling complex, multi-step attack chains, which are critical in real-world penetration testing scenarios. The consistent performance of VulnBot across diverse machines underscores its robustness and adaptability, making it a reliable tool for practical cybersecurity applications.

\subsection{Retrieval Augmented Generation (RQ4)}
\label{RAG}
To further investigate whether prior penetration knowledge can enhance the performance of our framework, we integrated the Memory Retriever module into the Llama3.1-405B model, which supports a 128k context window. This integration leverages RAG to improve the model's contextual understanding and task-specific optimization. In this experiment, we evaluated the performance of three distinct systems: Llama3.1-405B with RAG, GPT-4o with Manual, and Llama3.1-405B with Manual. The data for GPT-4o and Llama3.1 with Manual were obtained from \cite{isozaki2024towards}, where human operators utilized PentestGPT tools.
To augment the contextual knowledge of native LLMs, we incorporated content from cybersecurity resources such as HackTricks \cite{Hacktricks} and HackingArticles \cite{Articles}. This content was segmented into 750-word chunks, and the resulting embeddings were stored in the Milvus vector database \cite{Milvus} for efficient retrieval. This approach enables the system to dynamically retrieve relevant historical data and prior knowledge, thereby mitigating the hallucination problem often encountered with LLMs.

\begin{figure}[ht]
\centering
\includegraphics[width=\linewidth]{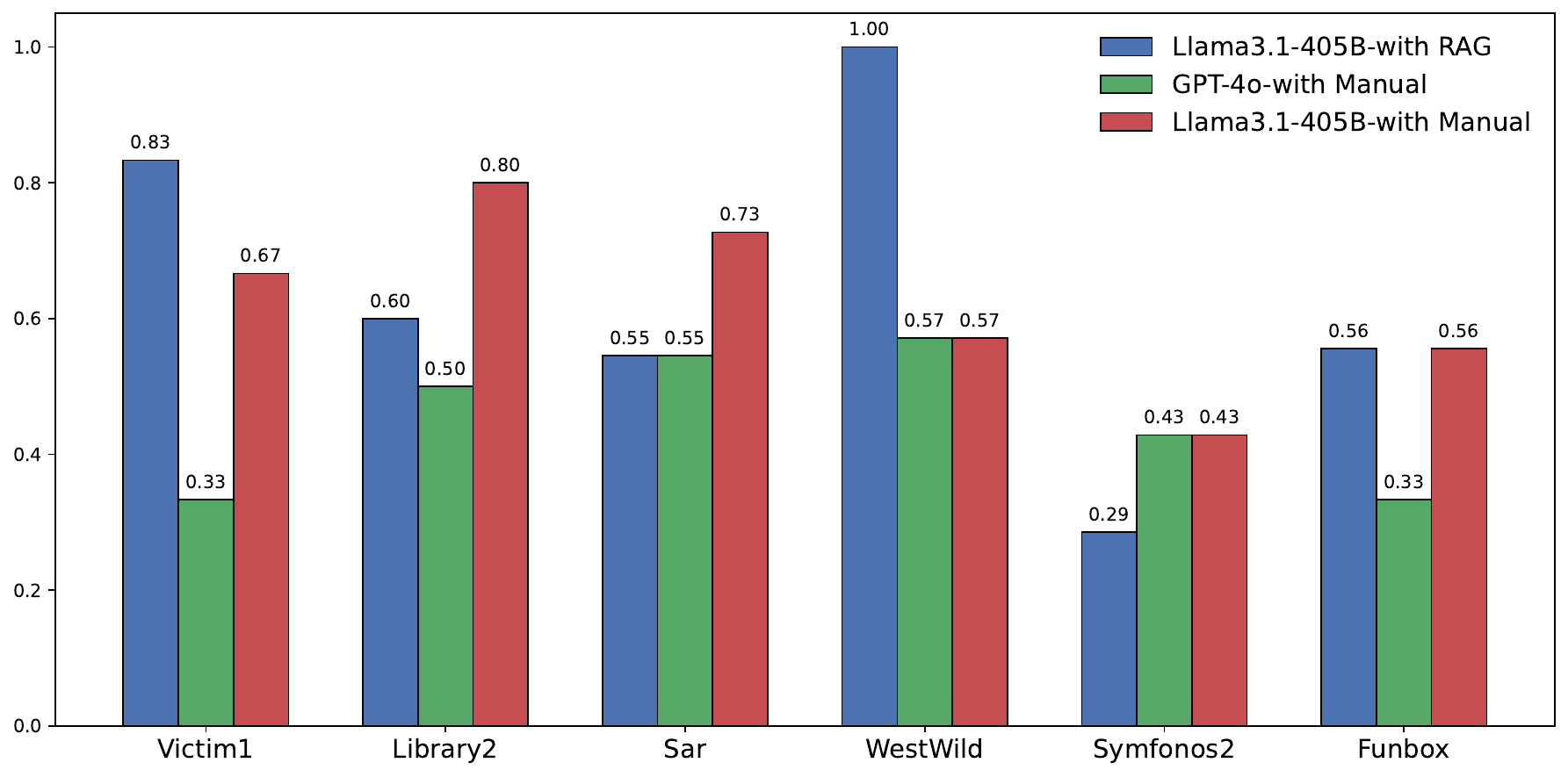}
\caption{Performance comparison of VulnBot with Memory Retriever module}
\label{fig:rag}
\end{figure}

Figure~\ref{fig:rag} illustrates the task completion rates of these models across six real-world machines. The results demonstrate that integrating the Memory Retriever module significantly enhances performance on specific machines, particularly Victim1 and WestWild. Notably, VulnBot successfully executed an end-to-end penetration of the WestWild machine, showcasing its ability to complete complex tasks autonomously. These findings highlight the advantages of retrieval-augmented approaches in improving the contextual understanding and task-specific optimization of penetration testing models. The integration of the Memory Retriever module not only enhances the model's ability to retrieve and utilize relevant information but also improves its overall performance in real-world penetration testing scenarios, achieving performance comparable to or even surpassing that of human operators.

%-------------------------------------------------------------------------------
\section{Discussion}
The results obtained highlight VulnBot's potential for efficient vulnerability detection and exploitation. However, our findings also reveal several challenges and areas for future research that need to be addressed to enhance its capabilities further.

\subsection{Limitations in Processing Non-Textual Information}
A significant limitation of VulnBot is its inability to process non-textual information, such as images or graphical interfaces generated by penetration testing tools. In real-world penetration testing scenarios, such is often critical for understanding attack surfaces and interpreting the results of security scans. Currently, VulnBot depends on manual descriptions to interpret these non-textual elements, which introduces a bottleneck in achieving full automation of the penetration testing process. Future iterations of VulnBot could address this limitation by incorporating image recognition and processing capabilities. This enhancement would enable the system to analyze and extract relevant information from screenshots and other graphical representations. 

\subsection{Real-World Performance and Challenges}
The real-world tasks in the AUTOPENBENCH include two CVEs from 2024. VulnBot completed one of these tasks using Llama3.3 and Llama3.1, despite both models having a knowledge cutoff in December 2023. This achievement highlights the reliability of our method, as it does not rely on prior knowledge of the vulnerabilities.
Despite these promising results in simulated environments, completing end-to-end penetration testing on real-world machines remains a significant challenge for VulnBot. Even with RAG to enhance the model’s contextual knowledge and task-specific optimizations,  VulnBot still faces difficulties in achieving full autonomy and success across all stages of a real-world penetration test. These challenges stem from the complexity of real-world systems, the dynamic nature of security vulnerabilities, and the need for precise execution of multi-step attack chains.

%-------------------------------------------------------------------------------
\section{Related Work}
\subsection{Vulnerability Detection and Exploitation}
Atropos introduces a novel fuzzing technique for detecting server-side vulnerabilities in PHP-based web applications. It utilizes snapshot-based, feedback-driven fuzzing, which is integrated directly with the PHP interpreter \cite{guler2024atropos}. Similarly, NAUTILUS focuses on identifying vulnerabilities in RESTful APIs through guided testing and parameter generation strategies, emphasizing complex API interactions and corner cases \cite{deng2023nautilus}. Furthermore, empirical studies, such as "Understanding Hackers’ Work", provide insights into the operational methods and challenges faced by offensive security practitioners, underscoring the need for improved tooling \cite{happe2023understanding}. 

Recent advancements in large language models (LLMs) have shown significant promise in vulnerability management. Liu et al. explore the use of ChatGPT for handling complex cybersecurity tasks, particularly bug report analysis. They introduce a self-heuristic prompt template that enhances ChatGPT's performance by summarizing domain knowledge from provided examples. This approach enables the model to learn task-specific characteristics, resulting in improved performance in bug report summarization through few-shot learning compared to zero-shot or general information prompts \cite{liu2024exploring}. Fang et al. further investigate LLMs' potential in exploiting one-day vulnerabilities, demonstrating that GPT-4 can autonomously exploit 87\% of a benchmark set of real-world vulnerabilities when provided with CVE descriptions. This research highlights the emergent capabilities of advanced LLMs in cybersecurity, while also raising concerns about their responsible deployment \cite{fang2024llm}.

\subsection{Automated Penetration Testing}
Recent research has explored integrating LLMs into penetration testing workflows, significantly enhancing automation and efficiency. Al-Sinani and Mitchel investigate the use of GPT-4 across various ethical hacking phases, including reconnaissance and post-exploitation \cite{al2024ai}. Tools like AUTOATTACKER \cite{xu2024autoattacker} and BreachSeek \cite{alshehri2024breachseek} leverage LLMs to automate post-breach activities and simulate cyberattacks, respectively, while CIPHER \cite{pratama2024cipher} specializes in assisting ethical researchers through structured augmentation methods. Frameworks such as PTGroup \cite{wu2024ptgroup} and HPTSA \cite{fang2024teams} demonstrate the potential of multi-agent systems and hierarchical planning in exploiting zero-day vulnerabilities. Furthermore, HackSynth \cite{muzsai2024hacksynth} and Pentest Copilot \cite{goyal2024hacking} highlight the role of crafted prompts and LLM integration in automating penetration testing sub-tasks. 

Happe and Cito further explore the application of LLMs in penetration testing, presenting use cases for both high-level task planning and low-level vulnerability hunting. Their work implements a feedback loop where LLM-generated actions are executed on a virtual machine via SSH, demonstrating the potential of LLMs to automate parts of penetration testing while raising ethical concerns about misuse \cite{happe2023getting}. Happe et al. also evaluated the ability of different LLMs to execute privilege escalation attacks in the simulation environment by developing a fully automatic tool Wintermute. The results show that GPT-4-turbo has achieved a remarkable success rate with the assistance of sufficient context information and state mechanism \cite{happe2024llms}. Huang and Zhu introduce PenHeal, a two-stage LLM framework for automated penetration and optimal remediation. Their system incorporates components like Planner, Executor, Estimator, Advisor, and Evaluator to streamline the process, using counterfactual prompting and a Group Knapsack Algorithm to prioritize effective and cost-efficient remediations \cite{huang2023penheal}.

However, these approaches face limitations, including dependency on detailed vulnerability descriptions (e.g., CVE data) for effective exploitation, instability and variability in performance across different tasks and environments, and the need for human intervention in complex or end-to-end penetration testing scenarios. Additionally, many systems struggle with long-term planning and adaptability in dynamic environments, and while assisted or multi-agent approaches improve success rates, fully autonomous agents still face challenges in achieving consistent and reliable results.
\subsection{Application of LLM in  Cybersecurity}
Beyond penetration testing, LLMs are increasingly applied to a broader range of cybersecurity tasks. Cycle enhances code generation capabilities through iterative self-refinement \cite{ding2024cycle}, while Guan et al. leverage LLMs to detect model optimization bugs in deep learning libraries \cite{guan2024large}. Fang et al. demonstrate the ability of LLMs to exploit recently disclosed vulnerabilities with high success rates. Tools like SecurityBot \cite{yan2024depending} integrate LLMs with reinforcement learning to improve cybersecurity operations, and AURORA automates the orchestration of APT attack campaigns \cite{wang2024sands}. Additionally, PTHelper \cite{de2024pthelper} streamlines penetration testing by integrating AI with state-of-the-art tools. These applications illustrate the versatility of LLMs in addressing diverse cybersecurity challenges.

%-------------------------------------------------------------------------------
\section{Conclusion}
In this paper, we present VulnBot, an autonomous penetration testing framework that LLMs and multi-agent systems. VulnBot is designed to emulate the collaborative workflows of human penetration testing teams, thereby addressing the inefficiencies and manual dependencies inherent in traditional penetration testing methodologies. By decomposing complex tasks into specialized phases—reconnaissance, scanning, and exploitation—and utilizing a PTG to ensure logical task execution, VulnBot demonstrates significant advancements in automating penetration testing workflows.

Our experimental results underscore VulnBot's superior performance relative to baseline models such as GPT-4 and Llama3. Incorporating RAG further augments VulnBot's capabilities, enabling it to execute end-to-end penetration tasks autonomously without human intervention. These findings highlight the potential of VulnBot to revolutionize the field of penetration testing by enhancing efficiency, scalability, and autonomy.
%-------------------------------------------------------------------------------

\bibliographystyle{plain}
\bibliography{usenix}

\appendix
\section{Prompt Example}
This appendix presents examples of the initial prompts used to guide the penetration tester through various stages of the cybersecurity training process. The two main types of prompts are the \textbf{Plan Session} prompt and the \textbf{Task Session} prompt.

\subsection{Plan Session Initial Prompt}
The Plan Session Initial Prompt is used to initiate a structured session, where the assistant helps guide the penetration tester through specific phases of the cybersecurity training process. This prompt defines the overall target and goal for the current phase while referencing the context of the previous phases.

\begin{center}
\begin{tcolorbox}[title=Plan Session Initialization, colback=white, colframe=black, arc=2mm, boxrule=0.5mm]
You are a \{name\} Assistant running on Kali Linux 2023. Your role is to assist testers in the cybersecurity training process. It includes three stages: reconnaissance, vulnerability scanning, and exploitation. You focus on the \{name\} phase. You must consider the context of the previous phases if applicable.\\

\textbf{Overall Target:}\\
\{init\_description\} \\

\textbf{Phase Goal:}\\
\{goal\} \\

\textbf{Optional Reference Tools (may or may not be used):}\\
\{tools\} \\

\textbf{The context of the previous phases:}\\
\{context\}
\end{tcolorbox}
\end{center}

\noindent
\textbf{Explanation of Variables:}
\begin{itemize}
    \item \texttt{\{name\}}: Refers to the specific assistant type (e.g., Reconnaissance Assistant, Exploitation Assistant, etc.).
    \item \texttt{\{init\_description\}}: A brief description of the overall target of the penetration testing process (e.g., \textit{I want to test <target IP>}).
    \item \texttt{\{goal\}}: Specifies the specific goal of the current phase (e.g., identify open ports in reconnaissance or exploit a vulnerability in exploitation).
    \item \texttt{\{tools\}}: Any reference tools that may assist in achieving the phase goal (e.g., Nmap for reconnaissance, Metasploit for exploitation).
    \item \texttt{\{context\}}: The results or insights gained from the previous phases, which may guide the approach for the current phase.
\end{itemize}

\noindent 
Note that \texttt{\{name\}}, \texttt{\{goal\}}, and \texttt{\{tools\}} are fixed attributes of the assistant character, while \texttt{\{context\}} is derived from the summarizer's output based on previous phases.

\subsection{Task Session Initial Prompt}
The Task Session Initial Prompt defines the assistant's behavior in managing individual tasks during the cybersecurity training process. It clarifies how tasks are described in detail and how task results should be evaluated.

\begin{center}
\begin{tcolorbox}[title=Task Session Initialization, colback=white, colframe=black, arc=2mm, boxrule=0.5mm]
You are a \{name\} Assistant running on Kali Linux 2023. Your role is to assist testers in the cybersecurity training process.\\

You will receive two types of input:
\begin{enumerate}
    \item \textbf{New Task:} When you receive a New Task, break it down into clear, actionable steps for the tester to follow.
    \item \textbf{Task Result:} When you receive a Task Result, verify if the task was successful based on the provided result.
\end{enumerate}

Reply with "yes" if you understood.
\end{tcolorbox}
\end{center}

%-------------------------------------------------------------------------------

%%%%%%%%%%%%%%%%%%%%%%%%%%%%%%%%%%%%%%%%%%%%%%%%%%%%%%%%%%%%%%%%%%%%%%%%%%%%%%%%
\end{document}